\documentstyle{article}
\begin{document} %\draft 25.08.2001, KL preprint KL TH 01/06

%----------------------------------------------------------

\begin{center}

{\LARGE \bf Many Faces of Born-Infeld Theory}
\vglue.2in

Sergei V. Ketov~\footnote{Supported in part by the 
`Deutsche Forschungsgemeinschaft' within the German 
national \newline ${~~~~~}$ programme `String Theory'.}
\vglue.1in

{\it Department of Physics, University of Kaiserslautern\\ 
    Erwin-Str\"odinger Str., 67653 Kaiserslautern, Germany}\\ 
\end{center}
\vglue.2in

\begin{abstract}
Born-Infeld theory is the non-linear generalization of Maxwell electrodynamics.
It naturally arises as the low-energy effective action of open strings, and it
is also part of the world-volume effective action of D-branes. The $N=1$ and 
$N=2$ supersymmetric generalizations of the Born-Infeld action are closely 
related to partial spontaneous breaking of rigid extended supersymmetry. We 
review some remarkable features of the Born-Infeld action and outline its 
supersymmetric generalizations in four dimensions. The non-abelian $N=1$ 
supersymmetric extension of the Born-Infeld theory and its $N=1$ 
supergravitational avatars are given in superspace.

\end{abstract}

% Underline for text or math

  \def\pp{{\mathchoice
            %{general format
               %[w] = length of horizontal bars
               %[t] = thickness of the lines
               %[h] = length of the vertical line
               %[s] = spacing around the symbol
              %
              %\kern [s] pt%
              %\raise 1pt
              %\vbox{\hrule width [w] pt height [t] pt depth0pt
              %      \kern -([h]/3) pt
              %      \hbox{\kern ([w]-[t])/2 pt
              %            \vrule width [t] pt height [h] pt depth0pt
              %            }
              %      \kern -([h]/3) pt
              %      \hrule width [w] pt height [t] pt depth0pt}%
              %      \kern [s] pt
          {%displaystyle
              \kern 1pt%
              \raise 1pt
              \vbox{\hrule width5pt height0.4pt depth0pt
                    \kern -2pt
                    \hbox{\kern 2.3pt
                          \vrule width0.4pt height6pt depth0pt
                          }
                    \kern -2pt
                    \hrule width5pt height0.4pt depth0pt}%
                    \kern 1pt
           }
            {%textstyle
              \kern 1pt%
              \raise 1pt
              \vbox{\hrule width4.3pt height0.4pt depth0pt
                    \kern -1.8pt
                    \hbox{\kern 1.95pt
                          \vrule width0.4pt height5.4pt depth0pt
                          }
                    \kern -1.8pt
                    \hrule width4.3pt height0.4pt depth0pt}%
                    \kern 1pt
            }
            {%scriptstyle
              \kern 0.5pt%
              \raise 1pt
              \vbox{\hrule width4.0pt height0.3pt depth0pt
                    \kern -1.9pt  %[e]=0.15pt
                    \hbox{\kern 1.85pt
                          \vrule width0.3pt height5.7pt depth0pt
                          }
                    \kern -1.9pt
                    \hrule width4.0pt height0.3pt depth0pt}%
                    \kern 0.5pt
            }
            {%scriptscriptstyle
              \kern 0.5pt%
              \raise 1pt
              \vbox{\hrule width3.6pt height0.3pt depth0pt
                    \kern -1.5pt
                    \hbox{\kern 1.65pt
                          \vrule width0.3pt height4.5pt depth0pt
                          }
                    \kern -1.5pt
                    \hrule width3.6pt height0.3pt depth0pt}%
                    \kern 0.5pt%}
            }
        }}

  \def\mm{{\mathchoice
                      %{general format %[w] = length of bars
                                       %[t] = thickness of bars
                                       %[g] = gap between bars
                                       %[s] = space around symbol
   %[w], [t], [s], [h]=3([g]) are taken from corresponding definitions of \pp
   %
                      %       \kern [s] pt
               %\raise 1pt    \vbox{\hrule width [w] pt height [t] pt depth0pt
               %                   \kern [g] pt
               %                   \hrule width [w] pt height[t] depth0pt}
               %              \kern [s] pt}
                  %
                       {%displaystyle
                             \kern 1pt
               \raise 1pt    \vbox{\hrule width5pt height0.4pt depth0pt
                                  \kern 2pt
                                  \hrule width5pt height0.4pt depth0pt}
                             \kern 1pt}
                       {%textstyle
                            \kern 1pt
               \raise 1pt \vbox{\hrule width4.3pt height0.4pt depth0pt
                                  \kern 1.8pt
                                  \hrule width4.3pt height0.4pt depth0pt}
                             \kern 1pt}
                       {%scriptstyle
                            \kern 0.5pt
               \raise 1pt
                            \vbox{\hrule width4.0pt height0.3pt depth0pt
                                  \kern 1.9pt
                                  \hrule width4.0pt height0.3pt depth0pt}
                            \kern 1pt}
                       {%scriptscriptstyle
                           \kern 0.5pt
             \raise 1pt  \vbox{\hrule width3.6pt height0.3pt depth0pt
                                  \kern 1.5pt
                                  \hrule width3.6pt height0.3pt depth0pt}
                           \kern 0.5pt}
                       }}

\catcode`@=11
\def\un#1{\relax\ifmmode\@@underline#1\else
        $\@@underline{\hbox{#1}}$\relax\fi}
\catcode`@=12

% Accents and foreign (in text):

\let\under=\unt                 % bar-under (but see \un above)
\let\ced=\ce                    % cedilla
\let\du=\du                     % dot-under
\let\um=\Hu                     % Hungarian umlaut
\let\sll=\lp                    % slashed (suppressed) l (Polish)
\let\Sll=\Lp                    % " L
\let\slo=\os                    % slashed o (Scandinavian)
\let\Slo=\Os                    % " O
\let\tie=\ta                    % tie-after (semicircle connecting two letters)
\let\br=\ub                     % breve
                % Also: \`        grave
                %       \'        acute
                %       \v        hacek (check)
                %       \^        circumflex (hat)
                %       \~        tilde (squiggle)
                %       \=        macron (bar-over)
                %       \.        dot (over)
                %       \"        umlaut (dieresis)
                %       \aa \AA   A-with-circle (Scandinavian)
                %       \ae \AE   ligature (Latin & Scandinavian)
                %       \oe \OE   " (French)
                %       \ss       es-zet (German sharp s)
                %       \$  \#  \&  \%  \pounds  {\it\&}  \dots

% Abbreviations for Greek letters

\def\a{\alpha}
\def\b{\beta}
\def\c{\chi}
\def\d{\delta}
\def\e{\epsilon}
\def\f{\phi}
\def\g{\gamma}
\def\h{\eta}
\def\i{\iota}
\def\j{\psi}
\def\k{\kappa}
\def\l{\lambda}
\def\m{\mu}
\def\n{\nu}
\def\o{\omega}
\def\p{\pi}
\def\q{\theta}
\def\r{\rho}
\def\s{\sigma}
\def\t{\tau}
\def\u{\upsilon}
\def\x{\xi}
\def\z{\zeta}
\def\D{\Delta}
\def\F{\Phi}
\def\G{\Gamma}
\def\J{\Psi}
\def\L{\Lambda}
\def\O{\Omega}
\def\P{\Pi}
\def\Q{\Theta}
\def\S{\Sigma}
\def\U{\Upsilon}
\def\X{\Xi}

% Varletters

\def\ve{\varepsilon}
\def\vf{\varphi}
\def\vr{\varrho}
\def\vs{\varsigma}
\def\vq{\vartheta}

% Calligraphic letters

\def\ca{{\cal A}}
\def\cb{{\cal B}}
\def\cc{{\cal C}}
\def\cd{{\cal D}}
\def\ce{{\cal E}}
\def\cf{{\cal F}}
\def\cg{{\cal G}}
\def\ch{{\cal H}}
\def\ci{{\cal I}}
\def\cj{{\cal J}}
\def\ck{{\cal K}}
\def\cl{{\cal L}}
\def\cm{{\cal M}}
\def\cn{{\cal N}}
\def\co{{\cal O}}
\def\cp{{\cal P}}
\def\cq{{\cal Q}}
\def\car{{\cal R}}
\def\cs{{\cal S}}
\def\ct{{\cal T}}
\def\cu{{\cal U}}
\def\cv{{\cal V}}
\def\cw{{\cal W}}
\def\cx{{\cal X}}
\def\cy{{\cal Y}}
\def\cz{{\cal Z}}

% Fonts

\def\Sc#1{{\hbox{\sc #1}}}      % script for single characters in equations
\def\Sf#1{{\hbox{\sf #1}}}      % sans serif for single characters in equations

                        % Also:  \rm      Roman (default for text)
                        %        \bf      boldface
                        %        \it      italic
                        %        \mit     math italic (default for equations)
                        %        \sl      slanted
                        %        \em      emphatic
                        %        \tt      typewriter
                        % and sizes:    \tiny
                        %               \scriptsize
                        %               \footnotesize
                        %               \small
                        %               \normalsize
                        %               \large
                        %               \Large
                        %               \LARGE
                        %               \huge
                        %               \Huge

% Math symbols

\def\slpa{\slash{\pa}}                            % slashed partial derivative
\def\slin{\SLLash{\in}}                                   % slashed in-sign
\def\bo{{\raise-.3ex\hbox{\large$\Box$}}}               % D'Alembertian
\def\cbo{\Sc [}                                         % curly "
\def\pa{\partial}                                       % curly d
\def\de{\nabla}                                         % del
\def\dell{\bigtriangledown}                             % hi ho the dairy-o
\def\su{\sum}                                           % summation
\def\pr{\prod}                                          % product
\def\iff{\leftrightarrow}                               % <-->
\def\conj{{\hbox{\large *}}}                            % complex conjugate
\def\ltap{\raisebox{-.4ex}{\rlap{$\sim$}} \raisebox{.4ex}{$<$}}   % < or ~
\def\gtap{\raisebox{-.4ex}{\rlap{$\sim$}} \raisebox{.4ex}{$>$}}   % > or ~
\def\TH{{\raise.2ex\hbox{$\displaystyle \bigodot$}\mskip-4.7mu \llap H \;}}
\def\face{{\raise.2ex\hbox{$\displaystyle \bigodot$}\mskip-2.2mu \llap {$\ddot
        \smile$}}}                                      % happy face
\def\dg{\sp\dagger}                                     % hermitian conjugate
\def\ddg{\sp\ddagger}                                   % double dagger
                        % Also:  \int  \oint              integral, contour
                        %        \hbar                    h bar
                        %        \infty                   infinity
                        %        \sqrt                    square root
                        %        \pm  \mp                 plus or minus
                        %        \cdot  \cdots            centered dot(s)
                        %        \oplus  \otimes          group theory
                        %        \equiv                   equivalence
                        %        \sim                     ~
                        %        \approx                  approximately =
                        %        \propto                  funny alpha
                        %        \ne                      not =
                        %        \le \ge                  < or = , > or =
                        %        \{  \}                   braces
                        %        \to  \gets               -> , <-
                        % and spaces:  \,  \:  \;  \quad  \qquad
                        %              \!                 (negative)

\font\tenex=cmex10 scaled 1200

% Math stuff with one argument

\def\sp#1{{}^{#1}}                              % superscript (unaligned)
\def\sb#1{{}_{#1}}                              % sub"
\def\oldsl#1{\rlap/#1}                          % poor slash
\def\slash#1{\rlap{\hbox{$\mskip 1 mu /$}}#1}      % good slash for lower case
\def\Slash#1{\rlap{\hbox{$\mskip 3 mu /$}}#1}      % " upper
\def\SLash#1{\rlap{\hbox{$\mskip 4.5 mu /$}}#1}    % " fat stuff (e.g., M)
\def\SLLash#1{\rlap{\hbox{$\mskip 6 mu /$}}#1}      % slash for no-in sign
\def\PMMM#1{\rlap{\hbox{$\mskip 2 mu | $}}#1}   %
\def\PMM#1{\rlap{\hbox{$\mskip 4 mu ~ \mid $}}#1}       %
\def\Tilde#1{\widetilde{#1}}                    % big tilde
\def\Hat#1{\widehat{#1}}                        % big hat
\def\Bar#1{\overline{#1}}                       % big bar
\def\sbar#1{\stackrel{*}{\Bar{#1}}}             % big bar with star
\def\bra#1{\left\langle #1\right|}              % < |
\def\ket#1{\left| #1\right\rangle}              % | >
\def\VEV#1{\left\langle #1\right\rangle}        % < >
\def\abs#1{\left| #1\right|}                    % | |
\def\leftrightarrowfill{$\mathsurround=0pt \mathord\leftarrow \mkern-6mu
        \cleaders\hbox{$\mkern-2mu \mathord- \mkern-2mu$}\hfill
        \mkern-6mu \mathord\rightarrow$}
\def\dvec#1{\vbox{\ialign{##\crcr
        \leftrightarrowfill\crcr\noalign{\kern-1pt\nointerlineskip}
        $\hfil\displaystyle{#1}\hfil$\crcr}}}           % <--> accent
\def\dt#1{{\buildrel {\hbox{\LARGE .}} \over {#1}}}     % dot-over for sp/sb
\def\dtt#1{{\buildrel \bullet \over {#1}}}              % alternate "
\def\der#1{{\pa \over \pa {#1}}}                % partial derivative
\def\fder#1{{\d \over \d {#1}}}                 % functional derivative
                % Also math accents:    \bar
                %                       \check
                %                       \hat
                %                       \tilde
                %                       \acute
                %                       \grave
                %                       \breve
                %                       \dot    (over)
                %                       \ddot   (umlaut)
                %                       \vec    (vector)

% Math stuff with more than one argument

\def\frac#1#2{{\textstyle{#1\over\vphantom2\smash{\raise.20ex
        \hbox{$\scriptstyle{#2}$}}}}}                   % fraction
\def\half{\frac12}                                        % 1/2
\def\sfrac#1#2{{\vphantom1\smash{\lower.5ex\hbox{\small$#1$}}\over
        \vphantom1\smash{\raise.4ex\hbox{\small$#2$}}}} % alternate fraction
\def\bfrac#1#2{{\vphantom1\smash{\lower.5ex\hbox{$#1$}}\over
        \vphantom1\smash{\raise.3ex\hbox{$#2$}}}}       % "
\def\afrac#1#2{{\vphantom1\smash{\lower.5ex\hbox{$#1$}}\over#2}}    % "
\def\partder#1#2{{\partial #1\over\partial #2}}   % partial derivative of
\def\parvar#1#2{{\d #1\over \d #2}}               % variation of
\def\secder#1#2#3{{\partial^2 #1\over\partial #2 \partial #3}}  % second "
\def\on#1#2{\mathop{\null#2}\limits^{#1}}               % arbitrary accent
\def\bvec#1{\on\leftarrow{#1}}                  % backward vector accent
\def\oover#1{\on\circ{#1}}                              % circle accent

\def\[{\lfloor{\hskip 0.35pt}\!\!\!\lceil}
\def\]{\rfloor{\hskip 0.35pt}\!\!\!\rceil}
\def\Lag{{\cal L}}
\def\du#1#2{_{#1}{}^{#2}}
\def\ud#1#2{^{#1}{}_{#2}}
\def\dud#1#2#3{_{#1}{}^{#2}{}_{#3}}
\def\udu#1#2#3{^{#1}{}_{#2}{}^{#3}}
\def\calD{{\cal D}}
\def\calM{{\cal M}}

\def\szet{{${\scriptstyle \b}$}}
\def\ulA{{\un A}}
\def\ulM{{\underline M}}
\def\cdm{{\Sc D}_{--}}
\def\cdp{{\Sc D}_{++}}
\def\vTheta{\check\Theta}
\def\fracm#1#2{\hbox{\large{${\frac{{#1}}{{#2}}}$}}}
\def\ha{{\fracmm12}}
\def\tr{{\rm tr}}
\def\Tr{{\rm Tr}}
\def\itrema{$\ddot{\scriptstyle 1}$}
\def\ula{{\underline a}} \def\ulb{{\underline b}} \def\ulc{{\underline c}}
\def\uld{{\underline d}} \def\ule{{\underline e}} \def\ulf{{\underline f}}
\def\ulg{{\underline g}}
\def\items#1{\\ \item{[#1]}}
\def\ul{\underline}
\def\un{\underline}
\def\fracmm#1#2{{{#1}\over{#2}}}
\def\footnotew#1{\footnote{\hsize=6.5in {#1}}}
\def\low#1{{\raise -3pt\hbox{${\hskip 0.75pt}\!_{#1}$}}}

\def\Dot#1{\buildrel{_{_{\hskip 0.01in}\bullet}}\over{#1}}
\def\dt#1{\Dot{#1}}
\def\DDot#1{\buildrel{_{_{\hskip 0.01in}\bullet\bullet}}\over{#1}}
\def\ddt#1{\DDot{#1}}

\def\Tilde#1{{\widetilde{#1}}\hskip 0.015in}
\def\Hat#1{\widehat{#1}}

% Aligned equations

\newskip\humongous \humongous=0pt plus 1000pt minus 1000pt
\def\caja{\mathsurround=0pt}
\def\eqalign#1{\,\vcenter{\openup2\jot \caja
        \ialign{\strut \hfil$\displaystyle{##}$&$
        \displaystyle{{}##}$\hfil\crcr#1\crcr}}\,}
\newif\ifdtup
\def\panorama{\global\dtuptrue \openup2\jot \caja
        \everycr{\noalign{\ifdtup \global\dtupfalse
        \vskip-\lineskiplimit \vskip\normallineskiplimit
        \else \penalty\interdisplaylinepenalty \fi}}}
\def\li#1{\panorama \tabskip=\humongous                         % eqalignno
        \halign to\displaywidth{\hfil$\displaystyle{##}$
        \tabskip=0pt&$\displaystyle{{}##}$\hfil
        \tabskip=\humongous&\llap{$##$}\tabskip=0pt
        \crcr#1\crcr}}
\def\eqalignnotwo#1{\panorama \tabskip=\humongous
        \halign to\displaywidth{\hfil$\displaystyle{##}$
        \tabskip=0pt&$\displaystyle{{}##}$
        \tabskip=0pt&$\displaystyle{{}##}$\hfil
        \tabskip=\humongous&\llap{$##$}\tabskip=0pt
        \crcr#1\crcr}}

% Journal abbreviations (preprints)

\def\pl#1#2#3{Phys.~Lett.~{\bf {#1}B} (19{#2}) #3}
\def\np#1#2#3{Nucl.~Phys.~{\bf B{#1}} (19{#2}) #3}
\def\prl#1#2#3{Phys.~Rev.~Lett.~{\bf #1} (19{#2}) #3}
\def\pr#1#2#3{Phys.~Rev.~{\bf D{#1}} (19{#2}) #3}
\def\cqg#1#2#3{Class.~and Quantum Grav.~{\bf {#1}} (19{#2}) #3}
\def\cmp#1#2#3{Commun.~Math.~Phys.~{\bf {#1}} (19{#2}) #3}
\def\jmp#1#2#3{J.~Math.~Phys.~{\bf {#1}} (19{#2}) #3}
\def\ap#1#2#3{Ann.~of Phys.~{\bf {#1}} (19{#2}) #3}
\def\prep#1#2#3{Phys.~Rep.~{\bf {#1}C} (19{#2}) #3}
\def\ptp#1#2#3{Progr.~Theor.~Phys.~{\bf {#1}} (19{#2}) #3}
\def\ijmp#1#2#3{Int.~J.~Mod.~Phys.~{\bf A{#1}} (19{#2}) #3}
\def\mpl#1#2#3{Mod.~Phys.~Lett.~{\bf A{#1}} (19{#2}) #3}
\def\nc#1#2#3{Nuovo Cim.~{\bf {#1}} (19{#2}) #3}
\def\ibid#1#2#3{{\it ibid.}~{\bf {#1}} (19{#2}) #3}

% Section heading and reference stuff

\def\sect#1{\bigskip\medskip \goodbreak \noindent{\bf {#1}} \nobreak \medskip}
\def\refs{\sect{References} \footnotesize \frenchspacing \parskip=0pt}
\def\Item{\par\hang\textindent}
\def\Itemitem{\par\indent \hangindent2\parindent \textindent}
\def\makelabel#1{\hfil #1}
\def\topic{\par\noindent \hangafter1 \hangindent20pt}
\def\Topic{\par\noindent \hangafter1 \hangindent60pt}

% ========================== END of def.tex ==========================

\section{Introduction}

~~~~The {\it Born-Infeld} (BI) non-linear electrodynamics \cite{bi} 
in Minkowski spacetime is defined by the Lagrangian
$$ \Lag\low{\rm BI}(F) = \fracmm{1}{b^2} \left\{ 
 1- \sqrt{ -\det (\h_{\m\n}+bF_{\m\n} )}\right\} ~,
\eqno(1)$$
where $F_{\m\n}=\pa_{\m}A_{\n}-\pa_{\n}A_{\m}$, $\m,\n=0,1,2,3$, 
and $b$ is the dimensional coupling constant ($b=1$ in what follows).
The BI theory implies famous taming of Coulomb self-energy of a point-like
electric charge \cite{bi}, while it shares with the Maxwell theory  
electric-magnetic self-duality \cite{bid} and physical propagation 
(no shock waves) \cite{caus}. In order to appreciate these highly non-trivial 
features, let's recall that a generic non-linear electrodynamics is defined by
 the field equations
$$ \eqalign{
\nabla\times \vec{E} = -\fracmm{\pa\vec{B}}{\pa t}~,\quad
\nabla \cdot \vec{B}=0~,\cr
\nabla\times \vec{H} = +\fracmm{\pa\vec{D}}{\pa t}~,\quad
\nabla \cdot \vec{D}=0~.\cr}\eqno(2)$$
If there exists a Lagrangian $\Lag(\vec{E},\vec{B})$, then we have 
$$ \vec{H}=-\fracmm{\pa\Lag}{\pa \vec{B}}\quad{\rm and}\quad
\vec{D}=+\fracmm{\pa\Lag}{\pa \vec{E}}~~.\eqno(3)$$

Lorentz invariance in four dimensions implies further restrictions,
$$ \Lag=\Lag(\a,\b)~,\quad{\rm where}\quad 
\a=\fracm{1}{2}(\vec{B}^2-\vec{E}^2)\quad
{\rm and}\quad \b=\vec{E}\cdot\vec{B}~.\eqno(4)$$
The electric-magnetic self-duality of the non-linear electrodynamics (2) under
rigid rotations,
$$ \vec{E} +i\vec{H} \to e^{i\q}(\vec{E} +i\vec{H}) \quad{\rm and}\quad
 \vec{D} +i\vec{B} \to e^{i\q}(\vec{D} +i\vec{B})~,\eqno(5)$$
together with eq.~(3) gives rise to a highly non-trivial non-linear constraint
\cite{bid},
$$ \vec{E}\cdot\vec{B}=\vec{D}\cdot\vec{H}~.\eqno(6)$$
In the manifestly Lorentz-covariant setting with $\Lag(F_{\m\n})$, it is 
natural to deal with the equations of motion and the Bianchi identities having
 the same form,
$$ \pa^{\n}\tilde{G}_{\m\n}=0\quad{\rm and}\quad 
 \pa^{\n}\tilde{F}_{\m\n}=0~,\eqno(7)$$
respectively, where we have defined
$$ \tilde{G}_{\m\n}(F)=\fracm{1}{2}\ve_{\m\n\l\r}G^{\l\r}(F)=
2\fracmm{\pa\Lag(F)}{\pa F^{\m\n}}~,\quad 
\tilde{F}_{\m\n}=\fracm{1}{2}\ve_{\m\n\l\r}F^{\l\r}~.\eqno(8)$$
Equation (6) then amounts to the non-linear constraint \cite{bid}
$$ G^{\m\n}\tilde{G}_{\m\n} + F^{\m\n}\tilde{F}_{\m\n}=0~.\eqno(9)$$

Causal propagation in a classical field theory follows from the dominant energy
condition on the energy-momentum tensor $T_{\m\n}$ (Hawking theorem) \cite{he}
$$  T_{00}\geq T_{\m\n}\quad {\rm for~all}~~\m~~{\rm and}~~{\n}~.\eqno(10)$$
It is straightforward (and very instructive) to verify that the BI theory (1) 
does satisfy both eqs.~(9) and (10). The absence of shock waves means that the
phase speed is phase-independent --- it is also the truly non-perturbative 
feature of the BI theory!

The BI theory possesses even more magical properties, such as the built-in 
upper bound for the electro-magnetic field-strength and the existence of exact
soliton-like solutions (called BIons) of finite total energy,
$\int d^3x\,T_{00} < \infty$ \cite{bion}.

The existence of the maximal electromagnetic field strength is obvious  from 
the form of the dual Hamiltonian density of the BI theory,
$$ \ch_{\rm dual}= 1-\sqrt{1-\vec{H}^2-\vec{E}^2+(\vec{H}\times\vec{E})^2}~~.
\eqno(11)$$  
In string theory, approaching the upper bound results in the breakdown of the 
BI theory due to a production of the open string massive states.

The BIon solution of electric charge $Q$ to the field equation
$$ \nabla\cdot\vec{D}=4\p Q\d(\vec{r})\eqno(12)$$
is given by
$$ \vec{D}=\fracmm{Q}{r^2}\,\vec{e}_r~,\qquad \vec{E}=\fracmm{\vec{D}}{\sqrt{
1+\vec{D}^2}}=\fracmm{Q}{\sqrt{r^4+Q^2}}\,\vec{e}_r~,\eqno(13)$$
so that the electric field singularity of the Maxwell theory is not present in
the BI theory, while the {\it effective} electric density $\r_{\rm eff.}$ 
of a point-like electric charge $Q$, 
$$\r_{\rm eff.}=\fracmm{1}{4\p}\nabla\cdot\vec{E}~,\eqno(14)$$
has a finite non-vanishing radius (of order $\sqrt{b}$).  

In string theory, when a {\it constant} Kalb-Ramond background $B_{\m\n}$ 
is turned on, $ F_{\m\n}\to F_{\m\n}+B_{\m\n}$, the BI theory in the limit 
$b=2\p\a'\to 0$ appears to be equivalent to a non-commutative $U(1)$ gauge 
field theory in flat spacetime with $\[x^{\m},x^{\n}\]=iB^{\m\n}$, via the 
Seiberg-Witten map \cite{swmap}. Note that Lorentz invariance is broken in
this case. The BI Lagrangian (1) in Euclidean spacetime interpolates between 
the Maxwell Lagrangian $\frac{1}{4}F^2$ for small $F$, and the topological 
density $\frac{1}{4}F\tilde{F}$ for large $F$, because of the identity
$$\sqrt{\det(F_{\m\n})}=\fracmm{1}{4}\abs{F\tilde{F}}~.\eqno(15)$$
The non-trivial (Euclidean) BI Lagrangian in the $b\to 0$ limit reads
$$ \fracmm{F^2}{\abs{F\tilde{F}}}~~,\eqno(16)$$
where we have used the relations
$$ \sqrt{\det(\ve^{1/2}+F)}\to \sqrt{\ve^2+\frac{\ve}{2}F^2
+\frac{1}{16}(F\tilde{F})^2}~\stackrel{\ve\to 0}{\longrightarrow}~ 
\fracm{1}{4}\abs{F\tilde{F}} +\ve\fracmm{F^2}{\abs{F\tilde{F}}}~.\eqno(17)$$
The first term on the r.h.s. of this equation is a total derivative, so that 
one arrives at eq.~(16) after rescaling eq.~(17) by a factor of $\ve^{-1}$.

The BI Lagrangian (1) in Euclidean spacetime obeys the (BPS) bound
$$\Lag_{\rm BI} = \sqrt{\left(1+\frac{1}{4}F\tilde{F}\right)^2
+\frac{1}{4}\left(F-\tilde{F}\right)^2}-1 \geq \frac{1}{4}\abs{F\tilde{F}}
\eqno(18)$$ 
that is saturated at {\it self-dual} field configurations, $F=\tilde{F}$, like
in the Maxwell case.

\section{BI theory and rigid supersymmetry}

The $N \,=\,$ 1 supersymmetric extension of the abelian BI action in four 
spacetime dimensions is the {\it Goldstone-Maxwell} (GM) action associated 
with {\it Partial} (1/2) {\it {Spontaneous Supersymmetry Breaking}} (PSSB) 
$N \,=\,$ 2 to $N \,=\,$ 1, whose Goldstone fields belong to a Maxwell (vector)
supermultiplet with respect to unbroken $N \,=\,$ 1 supersymmetry \cite{bg,tr}.

Manifest supersymmetry does not respect the standard determinantal form of the
 BI Lagrangian in eq.~(1). Moreover, eq.~(1) is not even the most elegant form 
of the BI theory! The complex bosonic variable, having the most 
natural $N \,=\,$ 1 supersymmetric extention (with linearly realized 
$N \,=\,$ 1 supersymmetry in superspace), is given by 
$$ \o=\a +i\b~,\qquad 
\a = \fracmm{1}{4}F^{\m\n}F_{\m\n}\equiv  \fracmm{1}{4}F^2~,
\quad \b = \fracmm{1}{4}F^{\m\n}\tilde{F}_{\m\n}\equiv  
\fracmm{1}{4}F\tilde{F}~.\eqno(19)$$
The BI Lagrangian (1) can be rewritten to the form $(b=1)$ 
$$ 
\Lag\low{\rm BI}(\o,\bar{\o})= 1- \sqrt{ 1+ (\o+\bar{\o})
+\frac{1}{4}(\o-\bar{\o})^2}~~,
\eqno(20)$$
or, equivalently, 
$$  
\Lag\low{\rm BI}(\o,\bar{\o})= \Lag\low{\rm ~free}+ \Lag\low{\rm 
~int.} \equiv  -\,\fracm{1}{2}\left(\o+\bar{\o}\right)
+\o\bar{\o}\cy(\o,\bar{\o})~,
\eqno(21)$$
whose structure function is given by
$$
\cy(\o,\bar{\o})\equiv \fracmm{1}{1+\fracmm{1}{2}(\o+\bar{\o})+
\sqrt{ 1+ (\o+\bar{\o})+\fracmm{1}{4}(\o-\bar{\o})^2}}~~.
\eqno(22)$$
The remarkably compact form of the BI action \cite{bg,tr}
$$  
\Lag\low{\rm BI}(\o,\bar{\o})= -\,\fracmm{1}{2}\left(\c+\bar{\c}
\right) =-{\rm Re}\,\c
\eqno(23)$$   
arises as the iterative solution to the simple non-linear constraint
$$ 
\c =-\,\fracm{1}{2}\c\bar{\c}+\o~.
\eqno(24)$$
This {\it Non-Linear Sigma-Model} (NLSM) form of the bosonic BI theory is 
quite natural from the viewpoint of PSSB \cite{book}. Indeed, to spontaneously
break any rigid symmetry, one may start with a free action that is invariant 
under the linearly realized symmetry, and then impose an invariant non-linear
  constraint that gives rise to the NLSM whose solutions break the symmetry.

The bosonic BI theory in the NLSM form is most convenient for a direct
supersymmetrization in superspace. One simply replaces the abelian bosonic 
field strength $F_{\m\n}$ by the abelian $N \,=\,$ 1 chiral spinor superfield
strength $W_{\a}$ obeying the standard off-shell $N \,=\,$ 1 superspace 
Bianchi identities, 
$$ 
\bar{D}_{\dt{\a}}W\low{\a}=0,\qquad D^{\a}W_{\a}-\bar{D}_{\dt{\a}}
\bar{W}^{\dt{\a}}=0~,\qquad \a=1,2~.
\eqno(25)$$ 
In the chiral basis the $N \,=\,$ 1 superfield $W_{\a}$ reads
$$ W^{\a}(x,\q)=\j^{\a}(x) +\q^{\b}\left[ (\s^{\m\n})\du{\b}{\a}F_{\m\n}(x)
+i\d\du{\b}{\a}D(x)\right] +\q^2i\pa^{\a\dt{\b}}\bar{\j}_{\dt{\b}}(x)~,
\eqno(26)$$
where $\j_{\a}(x)$ is the fermionic superpartner (Goldstone fermion) of the BI 
vector field, and $D$ is the real auxiliary field. The  $N \,=\,$ 1 
superextension of $\o$ is simply given by $\frac{1}{2}D^2W^2$, where
$W^2\equiv W^{\a}W_{\a}$ and $D^2=D^{\a}D_{\a}$. The $N \,=\,$ 1 
manifestly supersymmetric abelian BI action \cite{bg} in the NLSM form reads 
\cite{tr}
$$  S_{\rm 1BI}=\int d^4xd^2\q\,X +{\rm h.c.},
\eqno(27)$$
where the $N \,=\,$ 1 chiral superfield Lagrangian $X$ obeys the non-linear 
constraint 
$$ 
X = \ha X\bar{D}^2\bar{X}+\ha W^{\a}W_{\a}~~.
\eqno(28)$$
The iterative solution to eq.~(28) gives rise to the superfield 
action \cite{bg}
$$ 
S_{\rm 1BI} = \frac{1}{2}\left(\int d^4xd^2\q\,W^2+{\rm h.c.}\right) 
+ \int d^4xd^4\q\,\cy(\frac{1}{2} D^2W^2,\frac{1}{2}
\bar{D}^2\bar{W}^2)W^2\bar{W}^2 
\eqno(29)$$ 
with {\it the same} structure function (22) as in the bosonic BI case.
 
The NLSM form (27) and (28) of the $N \,=\,$ 1 BI action is also most useful 
in proving its invariance under the second (non-linearly realized and 
spontaneously broken) $N=1$ supersymmetry with the rigid spinor parameter 
$\h^{\a}$ \cite{bg,tr}.~\footnote{The spontaneously broken supercharges do not
 exist, but the supercurrents do.} 
$$ 
\d_2 X=\h^{\a}W_{\a}~,\quad \d_2W_{\a}=\h_{\a}\left(1-\ha\bar{D}^2
\bar{X} \right)+i\bar{\h}^{\dt{\a}}\pa_{\a\dt{\a}}X~~,
\eqno(30)$$
and its $N \,=\,$ 1 supersymmetric electric-magnetic self-duality as well. 
The latter amounts to a verification of the non-local constraint \cite{kuz}
$$ \int d^4x d^2\q (W^2+M^2)= \int d^4x d^2\bar{\q} (\bar{W}^2+\bar{M}^2
)~, \quad{\rm where}\quad  \fracmm{i}{2} M_{\a}=\fracmm{\d S_{\rm
1BI}}{\d W^{\a}}~,\eqno(31)$$
which is just the N=1 supersymmetric extension of eq.~(9).

The $N \,=\,$ 1 supersymmetric BI action can be put into the $N \,=\,$ 1 
{\it superconformal} form by inserting a conformal compensator ($N \,=\,$ 1
chiral superfield) $\F$ into the non-linear constraint (28) as follows  
\cite{kuz}:
$$ 
X = \fracmm{X}{2\F^2}\bar{D}^2\left(\fracmm{\bar{X}}{\bar{\F}^2}
\right) +\ha W^{\a}W_{\a}~~.
\eqno(32)$$ 
Equation (28) is recovered from eq.~(32) in the gauge $\F=1$. Varying the 
action (32) with respect to $\F$ yields $W_{\a}=0$ 
and, hence, $\o=0$ that, in turn, implies $F^2=F\tilde{F}=0$. Thus we arrive 
at the $N=1$ superconformal extension of the ultra-BI (conformal) bosonic
theory of Bialyncki-Birula \cite{pol}.

An off-shell $N \,=\,$ 2 supersymmetric Maxwell field strength also
 exists in $N \,=\,$ 2 superspace, where it is given by the $N \,=\,$ 2 
restricted chiral superfield $\cw$ subject to the $N \,=\,$ 2 Bianchi 
identities \cite{book}, 
$$ \bar{D}^{\dt{\a}}_i\cw=0 \quad{\rm and}\quad
 D_{ij}\cw=\bar{D}_{ij}\bar{\cw}~,\qquad i,j=1,2~.\eqno(33)$$
We use the notation
$$ D_{ij}=D^{\a}_iD_{j\a}~,\quad D^4=\prod_{i,\a}D^i_{\a}=
\fracm{1}{12}D_{ij}D^{ij}~,\eqno(34)$$
for the products of the $N \,=\,$ 2 flat superspace supercovariant derivatives,
and similarly for the $N \,=\,$ 2 superspace anticommuting coordinates 
$\q^{\a}_i$. In the chiral basis the $N \,=\,$ 2 superfield $\cw$ takes the
form
$$ \cw(x,\q)=\f +\q^{\a}_i\j^i_{\a}-\frac{1}{2}(\q\cdot\vec{\t}\q)
\vec{D} +\frac{i}{8}(\q\s^{\m\n}\q)F_{\m\n}-i(\q^3\cdot\pa\bar{\j})+\q^4\bo
\bar{\f}~,\eqno(35)$$
where $\f(x)$ is the complex scalar, $\j^i_{\a}(x)$ is the $SU(2)$
doublet of fermions, and $\vec{D}(x)$ is the auxiliary $SU(2)$
triplet. 

The  manifestly $N \,=\,$ 2 supersymmetric extension of the bosonic BI theory 
in the form (21) is given by
\cite{sk2} 
$$ S_{\rm 2BI} = \fracm{1}{2}\int d^4xd^4\q\,\cw^2 +\fracm{1}{4}\int d^4x
d^8\q\,\cy(K,\bar{K})\cw^2\bar{\cw}^2~,\eqno(36)$$
where we have introduced the $N \,=\,$ 2 chiral superfield extension of the 
bosonic $\o$ variable in eq.~(19),
$$ K= D^4\cw^2~,\eqno(37)$$
and used the same structure function $\cy$ of eq.~(22) 
as in the bosonic BI case.

The manifest $N \,=\,$ 2 supersymmetric extension of the BI-NLSM in eqs.~(23) 
and (24) was first proposed in ref.~\cite{sk2e},
$$  S_{\rm 2GM} = \fracm{1}{4}\int d^4xd^4\q\,\cx^2 +{\rm h.c.},\eqno(38)$$ 
where the $N \,=\,$ 2 chiral superfield Lagrangian $\cx$ obeys the non-linear
 $N \,=\,$ 2 superfield constraint \cite{sk2e}
$$ \cx =\fracm{1}{4}\cx\bar{D}^4\bar{\cx}+\cw^2~.\eqno(39)$$

Unlike the  $N \,=\,$ 1 BI actions (27) and (29), the $N \,=\,$ 2 actions (36)
 and (38) merely coincide modulo terms with the {\it spacetime} derivatives of
 $\cw$, i.e. \cite{sk2}
$$  S_{\rm 2GM} =  S_{\rm 2BI} +\co(\pa_{\m}\cw)~.\eqno(40)$$

The  $N \,=\,$ 2 supersymmetric extension of electric-magnetic self-duality in
 the $N \,=\,$ 2 non-linear electrodynamics characterized by some action 
$S(\cw,\bar{\cw})$ amounts to the relation \cite{kuz}
$$  \int d^4x d^4\q (\cw^2+\cm^2)= \int d^4x d^4\bar{\q} (\bar{\cw}^2
+\bar{\cm}^2)~,\quad \fracmm{i}{4}\cm=\fracmm{\d S}{\d\cw}~~, \eqno(41)$$
which is the natural generalization of eq.~(31). As was demonstrated in
ref.~\cite{kuz}, it is the action (38) that precisely obeys eq.~(41). Hence,
the higher-derivative terms present in the action (38) on the top of the 
action (36) should be taken seriously -- e.g., they contribute to the 
effective worldvolume (static-gauge) action of a D3-brane propagating 
in six dimensions. The leading terms of the action (38) up to the 8th order in 
$(\cw,\bar{\cw})$ were computed in ref.~\cite{bik},  
$$\eqalign{
S_{\rm 2GM} =  & \fracm{1}{4}\left(\int d^4xd^4\q\,\cw^2 +{\rm h.c.}\right)
+ \fracm{1}{8}\int d^4xd^8\q \left\{ \cw^2\bar{\cw}^2\left[ 
2+ D^4\cw^2+\bar{D}^4\bar{\cw}^2 \right]\right. \cr
  & -\fracm{1}{9}\cw^3\bo\bar{\cw}^3+\fracm{1}{2}\cw^2\bar{\cw}^2
\left[ (D^4\cw^2+\bar{D}^4\bar{\cw}^2)^2+D^4\cw^2\bar{D}^4\bar{\cw}^2\right]
\cr
 & \left. -\fracm{1}{6}D^4\cw^2\bar{\cw}^3\bo\cw^3-\fracm{1}{6}\bar{D}^4
 \bar{\cw}^2\cw^3\bo\bar{\cw}^3 +\fracm{1}{288}\cw^4\bo^2\bar{\cw}^4\right\}
+\ldots~,\cr}\eqno(42)$$
while they coincide with the ones obtained from the non-linear realization of 
 the PSSB $N \,=\,$ 4 to $N \,=\,$ 2 in four spacetime dimensions \cite{bik}. 

The  $N \,=\,$ 2 supersymmetric extension of the BI action can, therefore, be 
interpreted as the most essential (if not full) part of the Goldstone-Maxwell 
action associated with the PSSB $N \,=\,$ 4 to $N \,=\,$ 2 in four spacetime 
dimensions \cite{sk2e,bik}. The transformation laws of extra 
(spontaneously broken)  $N \,=\,$ 2 supersymmetry have the form \cite{sk2}
$$  \d\cx=2\L\cw~,\quad \d\cw=\L\left(1-\frac{1}{4}\bar{D}^4\bar{\cx}\right)
+\ldots~,\eqno(43)$$
where the dots stand for the higher-derivative terms, and the  $N \,=\,$ 2 
rigid supersymmetry parameter is given by
$$ \L(\q) =\l+\q^{\a}_i\l^i_{\a}+\q^{ij}\l_{ij}~.\eqno(44)$$   
Here $\l$ is the complex parameter of two broken translations, $\l^i_{\a}$ are
the Grassmann parameters of two broken supersymmetries, and $\l_{ij}$ are the
parameters of the spontaneously broken R-symmetry $SU(2)$. Accordingly, the
complex scalar $\f=P+iQ$ is the Goldstone scalar associated with two  
spontaneously broken
translations, whereas $\j^i_{\a}$ are two Goldstone spinors associated with 
 two spontaneously broken supersymmetries. 

Via a complicated field redefition, the leading bosonic part of the  
$N \,=\,$ 2 supersymmetric BI action can be rewritten to the form \cite{sk2e}
$$ S_{\rm brane} = -\int d^4x \sqrt{-\det(\h_{\m\n}+F_{\m\n}
+\pa_{\m}P\pa_{\n}P+\pa_{\m}Q\pa_{\n}Q)}~,\eqno(45)$$
which is characteristic to a D3-brane propagating in six spacetime dimensions.

\section{Non-abelian $N \,=\,$ 1 supersymmetric BI action}

The simple structure of the $N \,=\,$ 1 supersymmetric abelian BI action in 
eq.~(27) dictated by the Gaussian non-linear superfield constraint (28) allows
us to easily construct its {\it non-abelian} (NBI) generalization \cite{kno}
which may be relevant for the effective description of the D3-brane clusters 
(i.e. the `spacetime-filling' D3-branes on the top of each other). 

The $N \,=\,$ 1 {\it Super-Yang-Mills} (SYM) theory in $N \,=\,$ 1 superspace 
is defined by the standard off-shell constraints:
$$ 
\{\de\low{\a},\de\low{\b} \}= \{ \bar{\de}_{\dt{\a}},\bar{\de}_{
\dt{\b}} \}=0~, \quad \{ \de\low{\a},\bar{\de}_{\dt{\b}}
\}=-2i\de_{\a\dt{\b}}~~,
$$
$$ 
\[ \de\low{\a},\de_{\b\dt{\b}} \] =2i\ve\low{\a\b}\hat{
\bar{W}}_{\dt{\b}}~~, \quad \[ \bar{\de}_{\dt{\a}},\de_{\b\dt{\b}} \]
=2i\ve_{\dt{\a}\dt{\b}} \hat{W}\low{\b}~~,
\eqno(46)$$
in terms of the $N \,=\,$ 1 covariantly-chiral (Lie algebra-valued) gauge 
superfield strength $\hat{W}_{\a}=\hat{W}_{\a}^at_a$ obeying the Bianchi 
identities~\footnote{The Lie algebra generators $t_a$ obey the relations
$\[t_a,t_b\]=f_{abc}t_c$ and $\tr(t_at_b)=-2\d_{ab}$.} 
$$ 
\bar{\de}_{\dt{\a}}\hat{W}\low{\a}=0~,\quad \bar{\de}_{\dt{\a}}\hat{
\bar{W}}{}^{\dt{\a}}=\de^{\a}\hat{W}\low{\a}~~.
\eqno(47) $$

The natural $N \,=\,$ 1 supersymmetric NBI action is \cite{kno} 
$$  
S\low{\rm NBI} = \int d^4xd^2\q\,\tr~\hat{\F} + {\rm h.c.}~,
\eqno(48) $$
whose $N \,=\,$ 1 covarianlty chiral Lagrangian $\hat{\F}$ is subject to 
the `minimal' non-abelian generalization of the abelian non-linear
constraint (28),
$$  
\hat{\F} = \frac{1}{2} \hat{\F}\bar{\de}^2\hat{\bar{\F}}+\frac{1}{2} 
 \hat{W}^2~~.
\eqno(49)$$ 

The leading contribution to the NBI action (48) is the standard $N \,=\,$ 1 
SYM action in superspace,
$$ 
S_{\rm SYM}= \fracm{1}{2} \int d^4xd^2\q\,\tr~\hat{W}^2  + {\rm h.c.}
\eqno(50)$$
The next (quartic in the SYM field strength) correction in the Yang-Mills 
sector (in components) is given by the non-abelian Euler-Heisenberg term
 \cite{kno},
$$
\fracm{1}{4}\tr\left[ (F^2)^2 + (F\tilde{F})^2\right]~.
\eqno(51)$$ 

Our NBI action does not have ordering ambiguities that  are well known in the
bosonic NBI theory, because an iterative solution to the NBI constraint (49) 
also specifies the order of the non-abelian quantities.

The $N \,=\,$ 2 supersymmetric NBI action \cite{kno,za} has similar structure.

\section{Born-Infeld supergravity}

It is of interest to construct possible {\it gravitational} avatars of the 
BI action (see, e.g., ref.~\cite{deg} for the earlier discussion without 
supersymmetry). Requiring local supersymmetry implies more constraints on the
possible BI-type gravity actions that are to be non-linear in the 
spacetime curvature \cite{kg}.

The $N \,=\,$ 1 supergravity in four dimensions is most
naturally described in curved superspace $z^M=(x^m,\q^{\m}, \bar{
\q}_{\dt{\m}})$, $m=0,1,2,3$ and $\m=1,2$, where we now have to 
distinguish between curved $(M)$ and flat $(A)$ indices related
by a supervielbein $E\du{A}{M}$ and its inverse $E\du{M}{A}$ with 
$E={\rm Ber}(E\du{A}{M})\neq 0$  \cite{gbook}. The supervielbein 
$E\du{A}{M}$ and a superconnection $\O_A$ are most conveniently 
described by (super)  one-forms,
$$ 
E_A=E\du{A}{M}(z)\pa_M\quad {\rm and}\quad \O=dz^M\O_M(z)
= E^A\O_A~,
\eqno(52)$$
where $\O_A$ take their values in the Lorentz algebra,
$$ 
\O_A=\ha\O\du{A}{bc}(z)M_{bc}=\O\du{A}{\b\g}M_{\b\g}+
\O\du{A}{\dt{\b}\dt{\g}}\bar{M}_{\dt{\b}\dt{\g}}~,
\eqno(53)$$
and $M\low{bc}~\sim~(M\low{\b\g},M_{\dt{\b}\dt{\g}})$ are the Lorentz 
generators.  The curved superspace covariant derivatives
$$ 
\cd_A=\left(\cd_a,\cd_{\a},\bar{\cd}^{\dt{\a}}\right)=E_A+\O_A
\eqno(54)$$ 
obey the algebra
$$
\[ \cd_A,\cd_B\}=T\du{AB}{C}\cd_C+{\cal R}_{AB}~,
\eqno(55)$$
where the supertorsion $T\du{AB}{C}$ and the (Lorentz algebra-valued) 
supercurvature ${\cal R}_{AB}=\ha {\cal R}\du{AB}{cd}M_{cd}={\cal
R}\du{AB}{\b\g}M_{\b\g} + {\cal R}\du{AB}{\dt{\b} \dt{\g}} \bar{M}_{
\dt{\b}\dt{\g}}$ have been introduced. We assume that the latter satisfy
the standard (Wess-Zumino) off-shell superfield constraints \cite{gbook} 
defining the {\it minima\/l} $N \,=\,$ 1  supergravity in superspace. 
As a result of the constraints
and the Bianchi identities, all the superfield components of the supertorsion
and the  supercurvature appear to be merely dependent upon three
(constrained) supertorsion tensors: the complex (covariantly) chiral
scalar superfield ${\cal R}$, the real vector superfield $G_a$ and the
complex (covariantly)  chiral superfield $W_{\a\b\g}$ that is totally
symmetric with respect to its  spinor indices \cite{gbook}. The bosonic
superfield ${\cal R}$ has an  auxiliary  complex scalar $B$ as the
leading component, while it also contains the spacetime scalar curvature
as another bosonic field component. Similarly,  the bosonic vector
superfield  $G_a$ has the spacetime Ricci curvature amongst its field 
components. The fermionic superfield $W_{\a\b\g}$ has the gravitino
field strength as its leading component, while it also contains the 
spacetime Weyl tensor $C_{\a\b\g\d}$ (totally symmetric on its spinor
indices) as the fermionic field component.

The NBI superfield NLSM constraint (49) can be considered as the powerful tool 
converting {\it any} fundamental (input) chiral superfield Lagrangian 
$(\hat{W}^2)$ into the corresponding BI-type chiral Lagrangian $(\hat{\F})$. 
The supergravitational analogue of  the (covariantly chiral) $N \,=\,$ 1 SYM 
spinor superfield strength $W_{\a}^at_a$ is  given by the super-Weyl curvature 
tensor $W_{\a\b\g}M^{\b\g}$. This essentially amounts to replacing the  
Yang-Mills gauge group by the Lorentz group in the $N \,=\,$ 1 NBI action.  
The standard Weyl supergravity action \cite{gbook}
$$ S_{\rm W}= \int  d^4x d^2\q\,\ce\,\tr\,W^2 +{\rm h.c.}\eqno(56)$$ 
is now easily extended to the new {\it Born-Infeld-Weyl} (BIW) 
supergravity action \cite{kg}
$$ S_{\rm BIW}=\int   d^4x d^2\q\ce\,\tr\,\cf +{\rm h.c.}~,
\eqno(57)$$
whose covariantly chiral (Lorentz algebra-valued) Lagrangian $\cf$ is 
a solution to the non-linear superfield constraint
$$ 
\cf= \fracm{1}{2} \cf(\bar{\cd}^2-4{\cal R})\bar{\cf}+W^2~.
\eqno(58)$$
We have also introduced the chiral local denisty 
$$ \ce=-\fracmm{1}{4}{\cal R}^{-1}(\bar{\cd}^2-4{\cal R})E^{-1} \eqno(59)$$
in eqs.~(56) and (57).
  
The subleading correction to the Weyl supergravity action in the BIW 
theory (57) is given by
$$ 
S_{\rm BR}=\fracm{1}{2}\int  d^4x d^4\q E^{-1} \, {W^2}\low{\a\b\g}
\bar{W}^2_{\dt{\a}\dt{\b}\dt{\g}}~,
\eqno(60)$$ 
whose purely bosonic (gravitational) part is proportional to the  
 {\it Bel-Robinson} (BR) tensor squared \cite{br},
$$ 
T_{mnpq}= R_{mspt}R\du{n}{s}{}\du{q}{t}+R_{msqt}R\du{n}{s}{}\du{p}{t}
-\fracm{1}{2}g_{mn}R_{prst}R\du{q}{rst}~~.
\eqno(61)$$
In {\it four} dimensions the BR tensor (61) can be rewritten to the form 
$$  
T_{mnpq}=R_{mspt}R\du{n}{s}{}\du{q}{t} + \tilde{R}_{mspt}\tilde{R}
\du{n}{s}{}\du{q}{t}~,
\eqno(62)$$ 
where  $\tilde{R}_{mspt}$ is the dual curvature. Moreover, the four-dimensional
 BR tensor is also known to be totally symmetric and pairwise traceless 
\cite{br}. The BR tensor squared, $T^2_{mnpq}\,$, can be considered as the 
gravitational analogue of the gauge Euler-Heisenberg term (51).

Unfortunately, even the leading terms of the BIW action have higher 
derivatives. Unlike the gauge theory that is quadratic in the field strength,
the Einstein gravity action is linear in the curvature. 
The $N \,=\,$ 1 Einstein supergravity action 
$$ S_{\rm SG} = -\,\fracmm{3}{\k^2}\int d^8z E^{-1} 
\eqno(63)$$
does not contain the supercurvature at all. Here $\k$ is the
gravitational coupling constant of dimension of length.

In fact, {\it any} full superspace action containing a supercurvature (even
linearly) gives rise to the terms that are non-linear in the component
curvature. As an example, let's consider the simplest invariant (BI-Einstein) 
action \cite{kg}
$$ 
S_{\rm BIE} =\int d^8z E^{-1} (\L+{\cal R}) +{\rm h.c.}~,
\eqno(64)$$
where $\L$ is a non-vanishing constant. In components, eq.~(64)
gives rise to the following bosonic terms: 
$$ 
S_{\rm bos.} = -\,\fracm{1}{9} \int d^4x\sqrt{-g}(R+\fracm{1}{3}
B\bar{B}) (2\L+B+\bar{B})~,
\eqno(65)$$
where the auxiliary complex scalar field $B$ is the leading component of 
${\cal R}$. The algebraic $B$-equation of motion has a solution, 
$$ 
B=\bar{B}= -\frac{1}{3}\L \pm \sqrt{\frac{1}{9}\L^2 -R}~.
\eqno(66)$$
Being inserted back into the action (65), this yields   
$$ 
S_{\rm bos.} = -\fracm{4}{27}\int d^4x\sqrt{-g}\left\{\fracm{4}{3}
\L R + (\frac{1}{9}\L^2-R)\left(\frac{1}{3} \L\mp\sqrt{\frac{1}{9}
\L^2-R}\right)\right\}~.
\eqno(67)$$
This action is already of the BI type, while it also implies taming of
the spacetime scalar curvature from above,
$$ 
R\leq (\frac{1}{3}\L)^2~.
\eqno(68)$$

Having chosen the upper sign (minus) choice in eq.~(67), we can  adjust 
the free parameter $\L$ as
$$ 
\L = \left( \fracmm{3}{2\k}\right)^2~, 
\eqno(69)$$
so that the leading term (in the curvature) in the action (67) takes the 
standard (Einstein-Hilbert) form, $-\fracmm{1}{2\k^2}R~$.


\begin{thebibliography}{99}
\bibitem{bi} M. Born and L. Infeld, Proc. Roy. Soc. {\bf A144} (1934) 425
\bibitem{bid} M.K. Gaillard and B. Zumino, {\it Non-linear electromagnetic
 self-duality and Legendre transformations}, hep-th/9712103;\\
G.W. Gibbons and D.A. Rasheed, \np{454}{95}{185}
\bibitem{caus} J. Plebanski, {\it Nordita Lectures on Non-linear 
Electrodynamics}, Niels Bohr Institute preprint 1968 (unpublished);\\
G. Boillat, J. Math. Phys. {\bf 11} (1970) 941
\bibitem{he} S.W. Hawking and G.F.R. Ellis, {\it The Large Scale Structure
of Spacetime}, Cambridge University Press, 1973 
\bibitem{bion} G.W. Gibbons, Nucl.Phys. {\bf B514} (1998) 603
\bibitem{swmap} N. Seiberg and E. Witten, JHEP {\bf 09} (1999) 032
\bibitem{bg} J. Bagger and A.A. Galperin, \pr{55}{97}{1091}
\bibitem{tr} M. Rocek and A.A. Tseytlin, \pr{59}{99}{106001} 
\bibitem{book} S.V. Ketov, {\it Quantum Non-linear Sigma-Models},
Springer-Verlag, 2000
\bibitem{kuz} S.M. Kuzenko and S. Theisen, JHEP {\bf 03} (2000) 034;
 Fortschr. Phys. {\bf 49} (2001) 273
\bibitem{pol} I. Bialyncki-Birula, Non-linear electrodynamics: variations on
a theme of Born and Infeld, in {\it Quantum Theory of Fields and Particles},
edited by B. Jancerwicz and J. Lukierski, World Scientific, 1983, p.~31
\bibitem{sk2} S.V. Ketov, \mpl{14}{98}{501}, Class. and Quantum Grav. 
{\bf 17} (2000) L91
\bibitem{sk2e} S.V. Ketov, \np{553}{99}{250}
\bibitem{bik} S. Bellucci, E. Ivanov and S. Krivonos, 
{\it $N \,=\,$ 2 and $N \,=\,$ 4 supersymmetric Born-Infeld theories from 
non-linear realizations}, hep-th/0012236; and {\it Towards the complete 
$N \,=\,$ 2 superfield Born-Infeld action with partially broken 
$N \,=\,$ 4 supersymmetry}, hep-th/0101195
\bibitem{kno} S.V. Ketov, Phys. Lett. {\bf 491B} (2000) 207
\bibitem{za} A. Refolli, N. Terzi and D. Zanon, Phys. Lett. {\bf 486B} 
(2000) 337
\bibitem{deg} S. Deser and G.W. Gibbons, \cqg{15}{98}{L35}
\bibitem{kg} S.J. Gates, Jr., and S.V. Ketov, Class. and Quantum Grav. {\bf 18}
 (2001) 3561
\bibitem{gbook} S.J. Gates, Jr., M. Grisaru, M. Rocek and W. Siegel, 
{\it Superspace}, Benjamin/Cummings Publishers, 1981
\bibitem{br} I. Robinson, unpublished lecture at King's College (1958);\\
L. Bel, in `{\it Les Theories Relativistes de la Gravitation}', CNRS, Paris;\\
S. Deser, J.S. Franklin and D. Seminara, Class. and Quantum Grav. {\bf 16}
(1999) 2815.
\end{thebibliography}
\end{document}